\documentclass[runningheads]{llncs}
\usepackage{graphicx}
\usepackage{tikz}
\usepackage{lipsum}
\usepackage{physics}
\usepackage{braket, graphicx}
\usepackage{caption}
\usepackage{subcaption}
\begin{document}
\title{Quantum optimization of coherent chaotic systems: A case for buses of Kathmandu. }
\titlerunning{Quantum optimization of coherent chaotic systems}
\author{Kiran Adhikari\inst{1} \and
Christian Deppe\inst{2} \and
Aman Ganeju \inst{3} \and
Iva Kumari Lamichhane \inst{3} \and
Rohit Bhattarai \inst{3}  \and
Manghang Limbu \inst{3} \and
Nishma Bhattarai \inst{3}}
\authorrunning{K. Adhikari et al.}
\institute{Institute for Communications Engineering,
Technical University of Munich, Germany
 \and Institute for Communications Technology, Technische Universität Braunschweig, Germany \and
Tribhuvan University, Nepal}

\maketitle              % typeset the header of the contribution

\centerline{\textbf{In memory of Ning Cai}}
\begin{abstract}
 In this paper, we propose a novel quantum computing approach to solve the real-world problem of optimizing transportation in bustling Kathmandu city. The transportation system in Kathmandu is chaotic, with no central authority controlling the transportation. We leverage this chaotic feature in our quantum optimization procedure. The quantum chaos theory, Wigner-Dyson distribution, surfaced as the most effective bus spacing distribution for a bus driver to maximize their profit. We investigate the statistical properties of the buses with real-time GPS bus location data and optimize bus spacing and interval distribution around the 27 km circular ring road in Kathmandu. Using tools like quantum simulation, eigenvalue distributions, and output wave function analysis, we show that such optimal bus spacing distribution could be achieved.

\end{abstract}
\section{Introduction}
In the early 1980s, Benioff and Feynman introduced the concept of quantum computing, highlighting that quantum mechanical computers outperformed classical ones for specific tasks \cite{benioff1980computer}. Feynman suggested applying quantum mechanics to computational problems. As stated in \cite{feynman2018simulating}, it was suggested that standard quantum systems could simulate complex quantum systems to solve problems beyond the capabilities of classical computers. Although the realization of such a quantum simulator was unclear at the time, Feynman significantly influenced the development of quantum computing. Subsequently, the concept of a quantum Turing machine was introduced, and the existence of universal models based on quantum mechanics was theoretically established \cite{deutsch1985quantum}. In simple terms, quantum models can achieve the same computational capabilities as classical computers. The first quantum algorithm was presented in \cite{deutsch1992rapid}, which demonstrated exponential acceleration in quantum computation compared to classical computation. Following that, quantum algorithms were proposed in \cite{bernstein1993quantum} and \cite{simon1997power}, highlighting the advantages of quantum computing in solving specific problems. However, these problems were rather artificial, and their practical applicability is limited.

In 1994, Shor introduced a quantum algorithm (named after him today) designed for factorizing large integers \cite{shor1994algorithms}. Large prime factorization is known to be an NP-hard problem and is the foundation for the security of the RSA public key cryptosystem. Solving this problem using classical computers requires exponential time. However, the Shor algorithm demonstrated that quantum computing could achieve the same task in polynomial time, posing a potential threat to the security of RSA encryption by making it easier to decipher. Subsequently, Grover proposed a quantum searching algorithm that could efficiently locate a specific item in an unordered database. The Grover algorithm takes advantage of quantum parallelism, enabling it to check all the data simultaneously in each iteration. This feature significantly reduces the computational complexity of solving the search problem. The unique computation capabilities of Shor and Grover algorithms have since captured widespread attention and interest. One of the areas where quantum algorithms provide substantial improvements is the problem of optimization  \cite{durr2006quantum,jordan2005fast,harrow2009quantum,childs2017quantum,brandao2017quantum,van2020quantum,kerenidis2020quantum,van2020convex,chakrabarti2020quantum}. Another quantum algorithm that offers advantages over the classical algorithm is K nearest-neighbour clustering \cite{Viladomat_Jasso_2023}.

Typically, optimization problems involve finding the extremum (maximum or minimum) of an objective function, which could represent various quantities like cost, energy, profit, or time. In many cases, this function depends on a substantial number of parameters, making a brute-force approach impractical due to the need to search through a high-dimensional parameter space. Quantum computers operate in Hilbert space, and this space grows exponentially with the number of qubits ($2^N$). The concept behind quantum computing is to leverage this vast state space through quantum entanglement to enhance the efficiency of finding the optimal solution, ideally achieving exponential speed-up \cite{lloyd1996universal,abrams1997simulation,abrams1999quantum,aspuru2005simulated,harrow2009quantum}. However, upon closer examination, it becomes apparent that the speed-up for classical optimization problems is often quite modest \cite{denchev2016computational,albash2018demonstration,smolin2014classical}. In contrast, quantum speed-up becomes most advantageous for problems that are directly tied to the quantum-mechanical description of natural phenomena. An excellent example is finding the many-electron wave function of a molecular system. Classical computers struggle to solve such problems accurately when the number of electrons exceeds a few tens due to the exponential growth of Hilbert space with electron count. Quantum computers, with their large state space, can simulate chemical systems and compute their properties, including correlations and reaction rates. Of course, this is under the condition that the challenge of efficiently mapping the fermionic problem to the available qubit hardware is overcome.

Most of our currently best-known quantum algorithms exploit huge structures in the problem to solve it exponentially faster than classical computers \cite{aaronson2022much}. However, the history of physics has taught us that tools from physics are remarkably powerful even when there is little structure in the underlying system. A compelling exemplification of this principle is found in the domains of thermodynamics and statistical mechanics. In this paper, we look for quantum algorithms for a chaotic system where there is very little structure in the underlying system. 

In this paper, the term "chaos" is not related to the usual Lyapunov exponent from chaos theory. Rather, it is supposed to indicate that such complex systems can be modelled by random matrix theory. In quantum chaos theory, the spectrum of the random matrix is used as a diagnosis of quantum chaos \cite{guhr1998random}.  
While it is possible to optimize such complex systems using classical techniques such as the Monte Carlo algorithm, we are particularly interested in looking for potential quantum advantages. This is because random matrix theory, on which such systems are modelled, was first inspired by properties of nuclear quantum systems, which are better simulated by quantum computers. 

Random matrix theory presents itself as a promising domain to look for such a problem. Random-matrix theory deals with the statistical properties of large matrices with randomly distributed elements. The probability distribution of the matrices is taken as input, from which the correlation functions of eigenvalues and eigenvectors are derived as output. From the correlation functions, one then computes the physical properties of the system. In the context of systems being well modeled by the Random Matrix Theory, the bus system in Cuernavaca, Mexico \cite{krbalek2000statistical} in the late 1990s has become a canonical physical system \cite{krbalek2001headways,krbalek2002headway,krbalek2008inter,akemann2011oxford}. Moreover, since the publishing of the seminal paper, the associated phenomenon is still being rediscovered in different systems like cars on a motorway \cite{abul2007modeling} and subway trains in New York City \cite{jagannath2017random}. A similar phenomenon can be expected in the transportation system of Kathmandu. This was one of the primary motivations for exploring quantum optimization of the transport system.

With rapid urbanization, transportation supply and demand imbalances have led to common problems like pollution, traffic congestion, and energy waste in various countries. The best solution is prioritizing public transport and improving urban public transport systems. But the problem is that the urban transit network design problem (UTNDP), which is concerned with the determination of a set of routes with corresponding schedules for an urban public transport system, is an NP-Hard problem, which means that in realistic time frames, it is not possible to find an optimal verifiable solution \cite{fan2010metaheuristic}. The large number of possible solutions and computational expenses make it impractical to explore the entire solution space. Therefore, a lot of different methods have been devised to tackle the UTNDP. (For a more detailed overview of these methods, see \cite{fan2010metaheuristic} and references therein). 

The UTNDP can be subdivided into two major components, namely the transit routing problem and the transit scheduling problem \cite{chakroborty2003genetic}. Generally, the urban transit routing problem involves the development of efficient transit routes (e.g., bus routes) on an existing road network with predefined pickup/drop-off points (e.g., bus stops). On the other hand, the urban transit scheduling problem (UTSP) is charged with assigning the schedules for passenger-carrying vehicles. Traditionally, the two phases are usually implemented sequentially (or iteratively), with the routes determined in advance of the schedule. In this paper, we concentrate on the urban transit routing problem and present a novel approach by giving an algorithm to model the problem in a quantum mechanical framework to make it more amenable to further quantum processing.

% Freeman Dyson writes in the foreword to the Oxford Handbook of Random Matrix Theory, "One recent application of random matrix theory to real life seems to be missing from this book. Either it is here, and I missed it, or the editors of the book missed it. This is the work of two physicists from the Czech Republic, Milan Krbalek and Petr Seba, who studied the bus system of the city of Cuernavaca in Mexico. Unlike other metropolitan bus systems, this system is decentralized, with no central authority and no timetables.
% ....... 
% The Gaussian unitary ensemble gives the best approximation to uniform spacing that the bus drivers can achieve based on the limited information available to them.
% ..... It is remarkable that such large a public benefit results from such a simple optimization process...  When an expert on markets tells me that some piece of financial wizardry is sure to benefit mankind, I am inclined to believe that a Cuernavaca bus driver might do the job better." \cite{}

The structure of this paper is as follows. In section \ref{sec:RandomMatrixTheory}, we discuss the basics of Random Matrix Theory, and in section \ref{sec:cuernavaca}, we discuss the transport system of Cuernavaca, which Random Matrix Theory can model. In Section \ref{sec:Model}, we model the bus system of Kathmandu as a coherent chaotic system using the Dyson gas model. In section \ref{sec:optimization}, we suggest an algorithm for optimizing the route and also suggest its relation to signatures of quantum chaos. In section \ref{sec:outlook}, we give an outlook for future directions, while in section \ref{sec:conclusion}, we summarize our paper.

\section{Random matrix theory} 
\label{sec:RandomMatrixTheory}
 It is believed that the dynamics of general many-body quantum chaotic systems beyond a timescale $t_{Th}$, the so-called “Thouless time,” follow predictions from random matrix theory (RMT) \cite{PhysRevLett.39.1167} i.e., their late-time behavior resembles that of a random matrix chosen from an ensemble consistent with the system’s symmetries \cite{cotler2017chaos,Dyson1962brownian,bohigas1984characterization}.

%, for instance, time-reversal symmetry \cite{bohigas1984characterization} 
%If, in contrast, the underlying dynamics are non-chaotic, the energy levels will not follow random matrix statistics but rather will behave like independent random variables from a Poisson process \cite{berry1977level}. 
Random matrix theory is used to characterize complex quantum systems when there is limited knowledge about the Hamiltonian. The fundamental hypothesis is that the Hamiltonian can be treated as a random matrix drawn from an ensemble with appropriate symmetries \cite{wigner1951statistical}. This approach is particularly useful for systems with many degrees of freedom and unknown interaction couplings among them.

%The conjecture, first mentioned in the work of Dyson \cite{Dyson1962statistical} that the statistical properties of energy levels of chaotic quantum systems may be described in terms of the theory of random matrices is widely accepted in various fields of physics. 

\begin{figure}[h]
    \centering
    \includegraphics[width=0.7\textwidth]{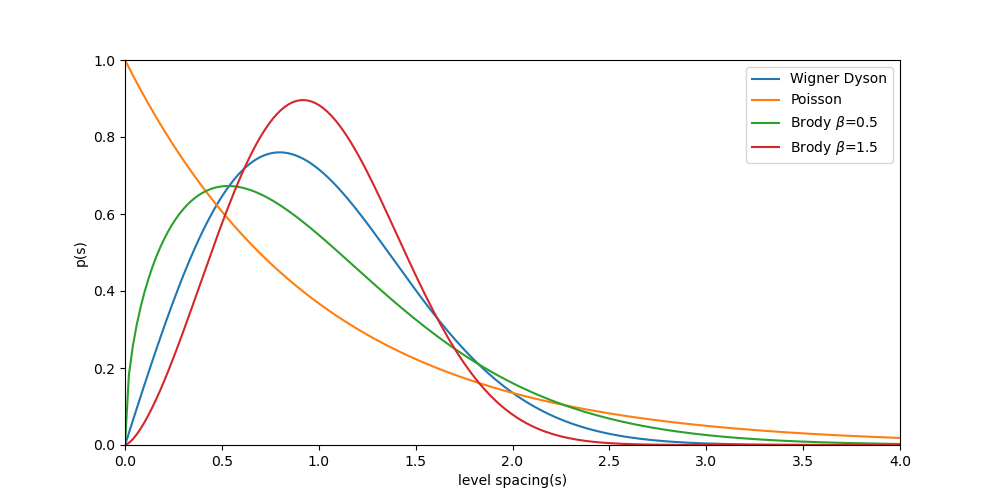} 
    \caption{level spacing distribution $P(s)$ for various systems}
    \label{fig:Distributions}
\end{figure}

Within this approach, one is often interested in the eigenspectrum of the Hamiltonian. Let the eigenvalues be $\lambda_1 \leq \lambda_2 \leq .... \leq \lambda_n $ and let $S_n = \lambda_{n+1} - \lambda_n$ be the consecutive splittings. The average value $D = \langle S_n \rangle$ is the mean splitting over the eigenvalue intervals. In integrable Hamiltonian models, the eigenvalues are uncorrelated; they are not prohibited from crossing and are usually described by Poisson statistics \cite{berry1977level}. In this case, the distribution $P(s)$ of the neighbouring spacing $s = S/D$, where $S$ is a particular spacing and $D$ is the mean distance between neighbouring intervals, is given by:
\begin{equation}
    P(s)  = \frac{1}{D}e^{-s}
\end{equation}
In contrast, the level spacing distribution P(s) of chaotic models is closely approximated by the Wigner-Dyson (WD) distribution:
\begin{equation}
    P(s) = b_\beta s^\beta e^{-a_\beta s^2}
\end{equation}
where contacts $b_\beta$ and $a_\beta$ are determined by normalization and $\beta$ depends on which universality class of random matrices the chaotic Hamiltonian belongs to \cite{Scharf1990Mar}: $\beta = 1$ for the Gaussian Orthogonal Ensemble (GOE), $\beta = 2$ for the Gaussian Unitary Ensemble (GUE), and $\beta = 4$ for the Gaussian symplectic ensemble (GSE), and $b_1 = \pi/2, a_1 = \pi /4$; $b_2 = 32/\pi^2, a_2 = 4/ \pi$; $b_4 = 262144/729 \pi^3, a_4 = 64/9 \pi$. Different distribution functions have been proposed to compare the spectral statistics with regular and chaotic limits and exhibit interpolation between them. For example, for the Gaussian Orthogonal Ensemble (GOE) statistics, one popular intermediate distribution is the Brody distribution \cite{brody1973statistical}:
\begin{align}
    P(s) = b(1+q)s^qe^{-bs^{q+1}}, b =\left[ \Gamma \left( \frac{2 + q}{1 + q} \right) \right] ^{q+1}
\end{align}
where $q = 0$ corresponds to the Poisson limit while $q = 1$ corresponds to Wigner-Dyson limit. Such distributions are shown in figure \ref{fig:Distributions}.

The Bohigas-Giannoni-Schmit (BGS) conjecture \cite{bohigas1984characterization}, \cite{guhr1998random} associates the quantum chaotic properties of a system with the correlations between its energy levels. Chaotic Hamiltonian exhibits level correlations in agreement with the predictions of random matrix theory (RMT) \cite{mehta2004random}. Adjacent eigenvalues show level repulsion and signals of spectral rigidity at larger energy scales. This understanding can be extended to the eigenspectrum of the Hamiltonian, allowing us to determine the chaotic nature of a quantum state based on its level spacing distribution \cite{chen2018universal}.

We will study one of the public bus systems of Kathmandu, namely, the Ring Road Bus system, and explore the chaotic nature of the system using various quantum chaos diagnostics as a function of eigenvalue level spacing and wave-function analysis. 

\subsection{Cuernavaca bus system and Dyson gas}
\label{sec:cuernavaca}

Dyson gas model is a mathematical model proposed by Freeman Dyson to describe the distribution of eigenvalues of random matrices in which Dyson gas is mirrored as an interacting many-particle system moving in one dimension \cite{guhr1998random}.

Buses of Cuernavaca are analogous to Dyson gas particles. Thus, the Dyson gas model can describe the statistical properties like bus spacing distribution of transport in Cuernavaca \cite{krbalek2000statistical}. Since very few changes in the initial conditions can bring out unpredictable changes in the Cuernavaca bus system as well as Dyson gas behaviours, both behave like chaotic quantum systems. This kind of complex system is characterized by minimum information content about the Hamiltonian, as a result of which their statistical properties follow random matrix ensembles( Gaussian Unitary Ensemble) prediction. 

The motion of buses, as well as Dyson gas particles, is along one dimension where $x_1, x_2,x_3,...,x_n$ denotes the position on the circle. Both systems are interactive, where the interaction between buses is indirectly facilitated by people assigned at different bus stops, which provide information about distance and time gap with the preceding bus ($x_1-x_2)$ and following bus ($x_2-x_3$). Likewise, the interaction potential that exists between buses in Cuernavaca is the function of the spacing between the buses \cite{krbalek2000statistical}:

\begin{equation}
    V(x)\approx1/\abs{x}^a
\end{equation}
where a is a positive constant.
In a similar sense, the interaction potential between Dyson gas particles is also a function of the distance between them ($\lambda$), which can be expressed as:
 \begin{equation}
    V(\lambda)\approx 1/ \lambda^a 
 \end{equation}
where $a$ is a positive constant.

\section{Modeling buses of Kathmandu as a coherent chaotic system}
\label{sec:Model}
It is well known that a probability distribution of the distances between subsequent buses with a centralized bus system is close to the Poisson distribution and can be described by the standard bus route model \cite{o1998jamming}. The origin of Poisson distribution in the centralized bus system comes from the fact that two neighbouring buses are not interacting with each other. In contrast to the centralized bus system in other cities, there is no centralized bus company in Kathmandu. Drivers work independently and try to optimize the amount of money earned while following a given route. Consequently, constraints such as a timetable that represents external influence on transport do not exist. This leads to competition among the drivers and to their mutual interaction which gives a departure from Poisson-like distribution.
 A Poisson-like distribution implies that the probability of close encounters between two buses is high (bus clustering), which conflicts with the effort of the drivers to maximize the number of transported passengers and, accordingly
to maximize the distance to the preceding bus. 
This, on the other hand, means that they tend not to arrive at a given bus stop immediately after their predecessors. To achieve that, they learn information about the time of departure of the previous bus from collaborators stationed at the bus stop where they arrive. Knowing the time interval to the preceding bus, the driver tries to optimize the distance to it by either slowing down or speeding up. In such a way, the obtained information leads to a direct interaction between buses and changes the statistical properties. 

The GUE properties of the bus arrival statistics can be well understood when regarding
the buses as a one-dimensional interacting gas. Buses here resemble Dyson gas particle behaviour, and buses are one-dimensional gas particles interacting with inverse square potential, i.e. if $x$ is the distance between the particles, then $1/x^2$ is their interaction potential. It has already been mentioned that the exact GUE statistics are obtained for a Coulomb interaction between the gas particles, i.e., for the
interaction potential $V$ given by:
\begin{equation}
\label{eq:Dyson_gas1}
    V = - \sum_{i < j} \log (|x_i - x_j|) + \frac{1}{2}\sum_i x_i^2
\end{equation}
In \eqref{eq:Dyson_gas1}, the second term represents a force confining the gas close to the origin. It is possible to consider more complicated Hamiltonian,
\begin{equation}
    H = \sum_{i = 1}^N \frac{m_i v_i^2}{2} + \sum_{i = 1}^N U(| x_{i-1} - x_i - l_i|)
\end{equation}
with $x_i$, $l_i$, $m_i$, and $v_i$ being the location, length, mass, and velocity of the ith vehicle.

Other factors of the Kathmandu transportation system can also be considered for realistic optimization. In some busy bus stops, there are traffic policemen with some probability to check the stay time of the bus so that they can minimize traffic jams. This can act as a cutoff for the stay time at those bus stops. Drivers also communicate with buses returning in opposite lanes to get information about the position of other buses. This can be modelled by considering a probabilistic interaction among buses that are not neighbours. 
Every driver usually encounters the widespread phenomenon of so-called bottleneck jam leading to transient clustering. So, one bus leaving earlier may be caught by another bus behind. Furthermore, at a particular length interval, there might be buses from different routes. In addition to the intra-route competition, in this case, there will be competition among buses from different routes, which we call inter-route interaction. All of these extra features can be considered for realistic optimization.

\subsection{Ring Road route}
In this paper, we will only consider one particular route of Kathmandu called the Ring Road route. The circumference of the Ring Road route is 27 km with a width of 31m on either side from the centre of the road and was constructed in 1977 with China's assistance \cite{shrestha2013analysis}. 

The ring road of Kathmandu, as its name suggests, is a complete ring-like transport system where the motion of vehicles is periodic along a closed path around the cities of Kathmandu and Lalitpur. This closed circular path gives a periodic boundary condition which makes it easier to do numerical and quantum simulations. Furthermore, we can now think of buses as a one-dimensional gas moving on a circle instead of a line, and we can safely ignore the second term of Eq \eqref{eq:Dyson_gas1}.

% \textcolor{red}{Include the Map of Ring road}
\begin{figure}[h]
    \centering
    \includegraphics[width=0.8\textwidth]{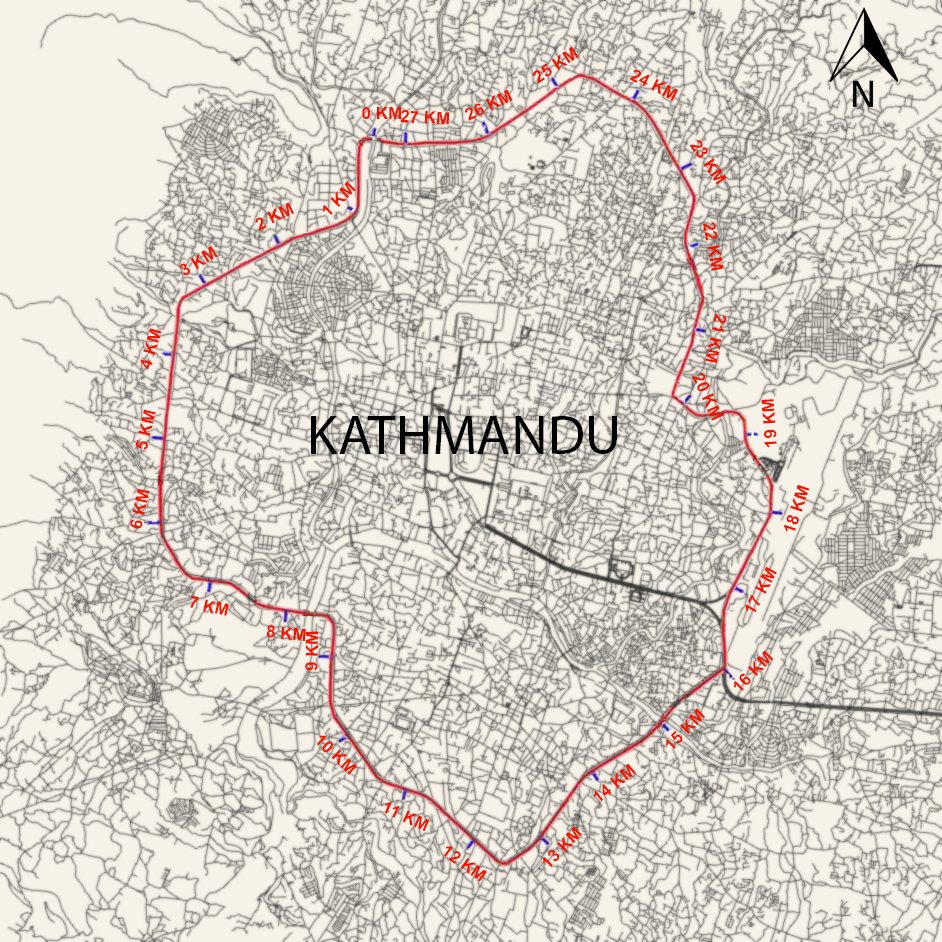}
    \caption{Map of Ring Road: The circumference of the Ring Road route is 27 km with a width of 31m on either side from the centre of the road.}
\end{figure}

\begin{figure*}
     \centering
     \begin{subfigure}[b]{0.4\textwidth}
         \centering
         \includegraphics[width =  \textwidth]{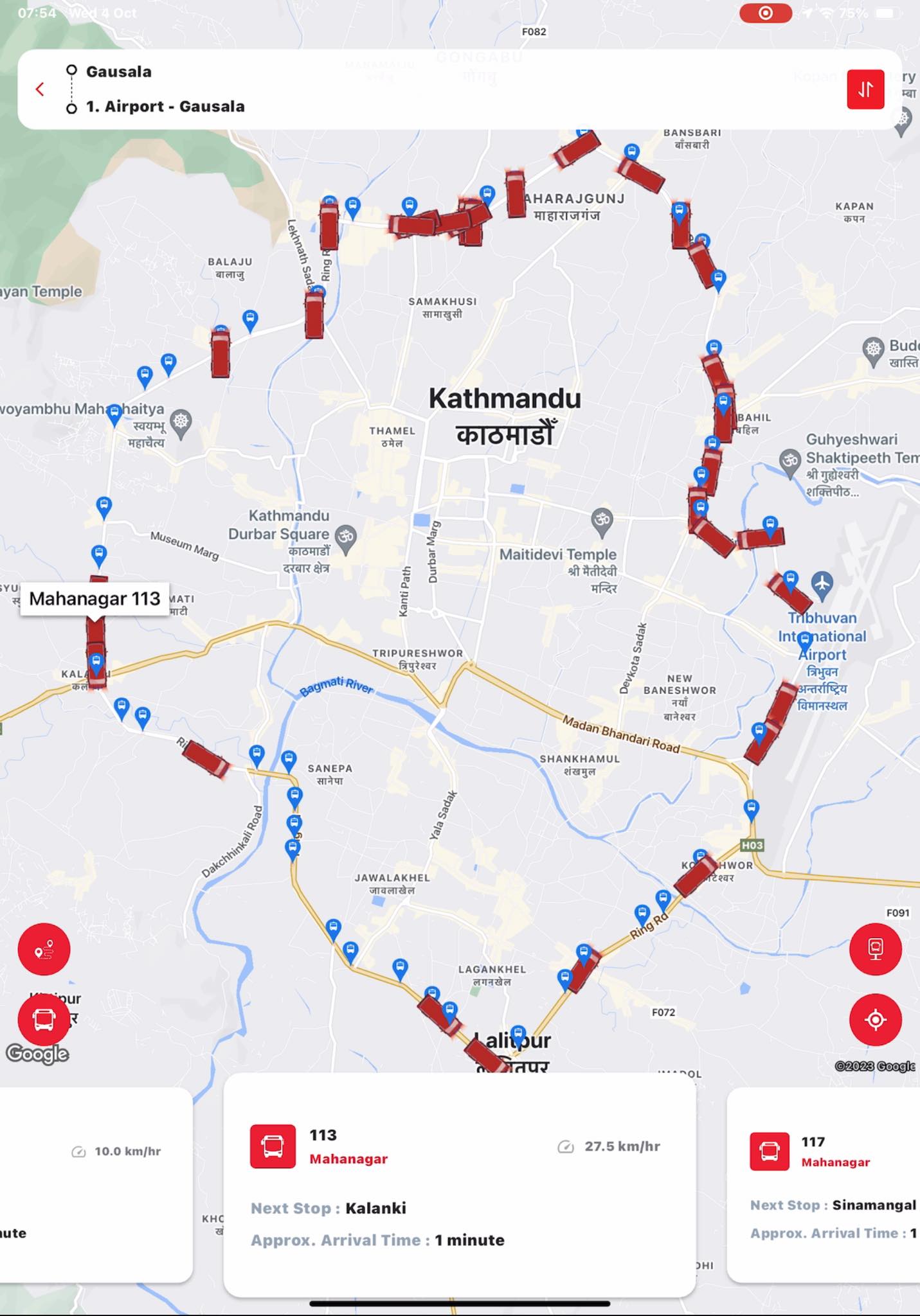}
         \caption{Clustering At 7:54 A.M}
        % \label{fig:y equals x}
     \end{subfigure}
     \hfill
     \begin{subfigure}[b]{0.4\textwidth}
         \centering
         \includegraphics[width = \textwidth]{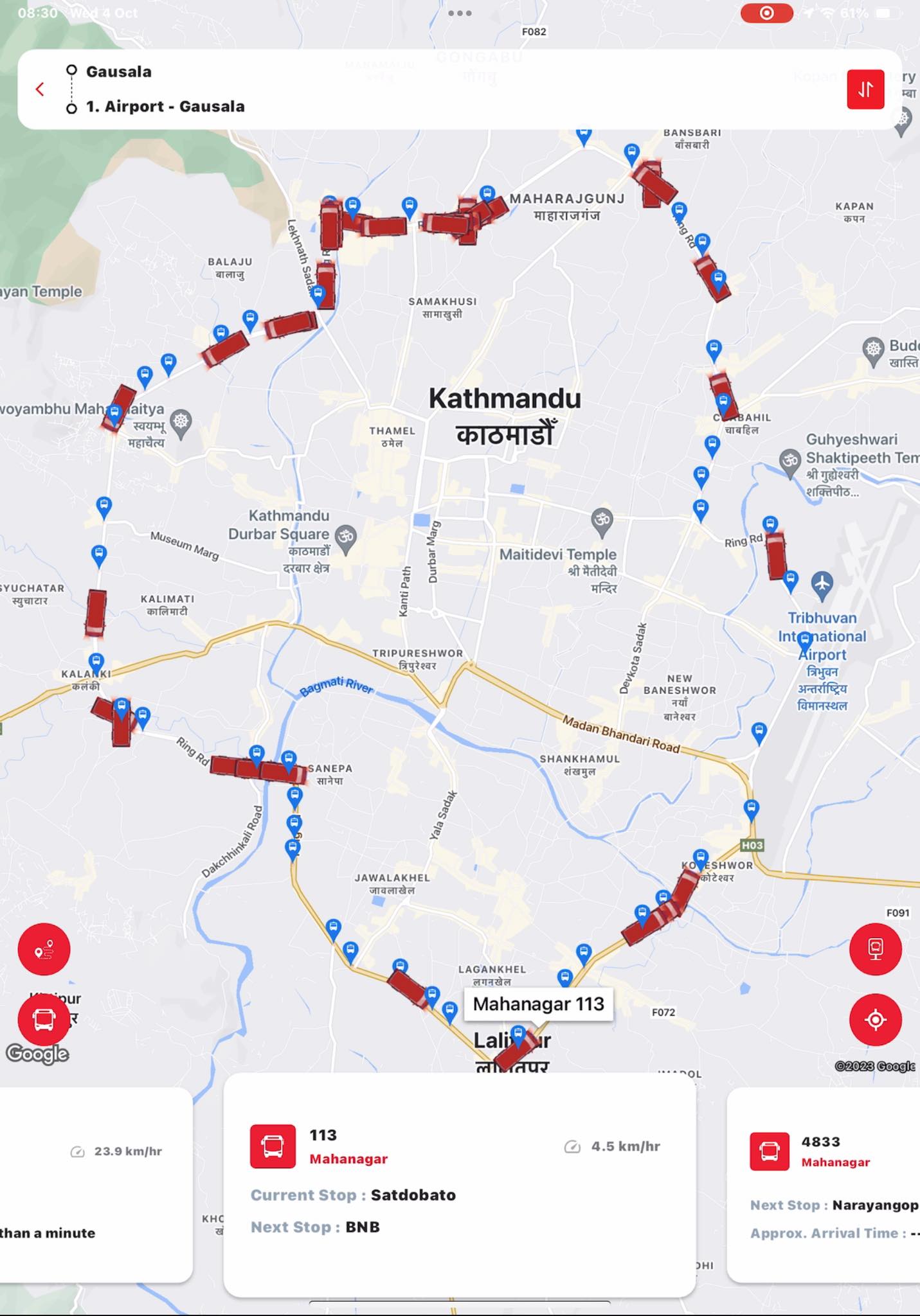}
         \caption{Clustering At 8:30 A.M}
        % \label{fig:three sin x}
     \end{subfigure}
          \hfill
     \begin{subfigure}[b]{0.4\textwidth}
         \centering
         \includegraphics[width = \textwidth]{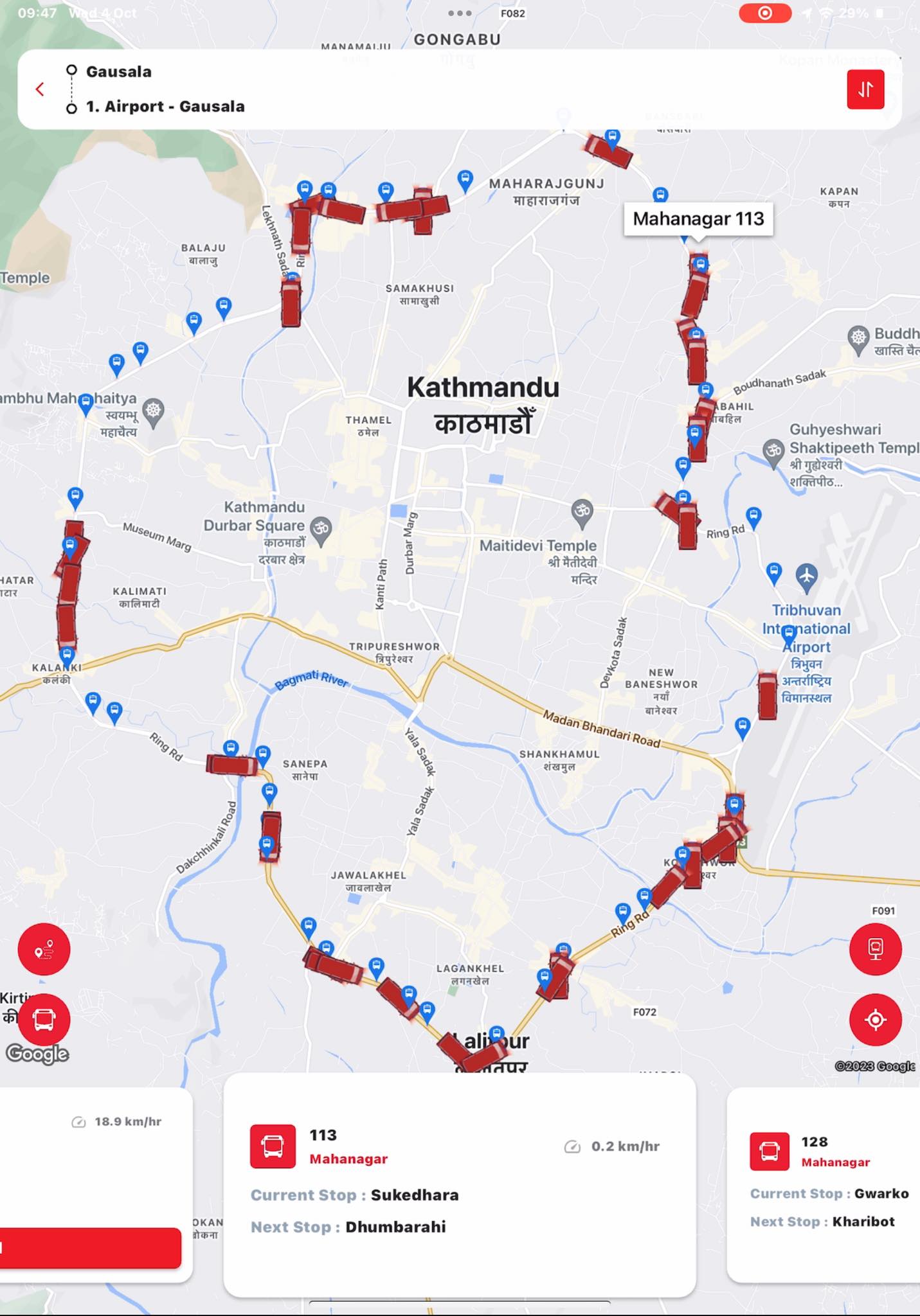}
         \caption{Clustering At 9:47 A.M}
         %\label{fig:three sin x}
     \end{subfigure}
    \hfill
    \begin{subfigure}[b]{0.4\textwidth}
         \centering
         \includegraphics[width =  \textwidth]{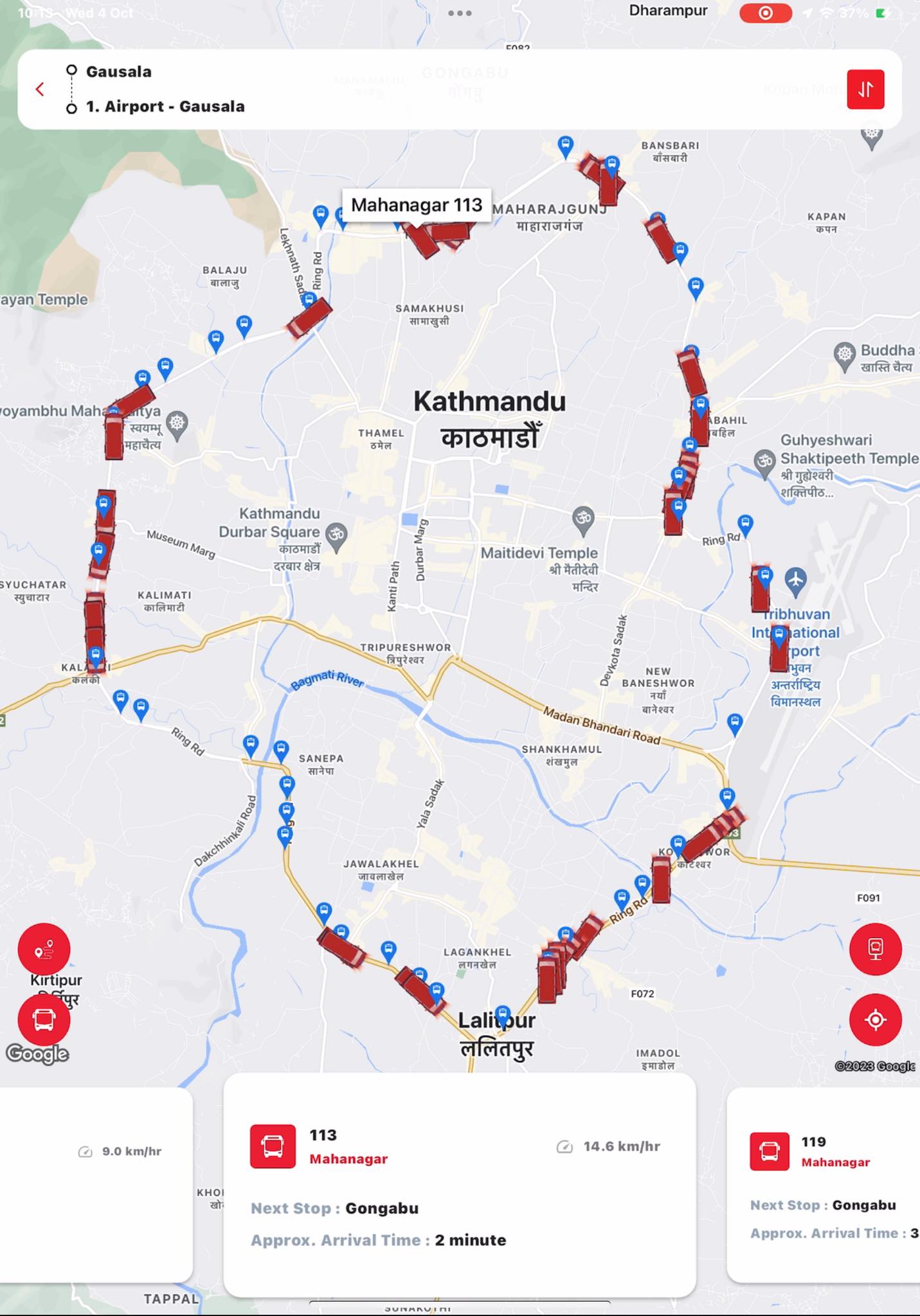}
         \caption{Clustering At 10:13 A.M}
        % \label{fig:y equals x}
     \end{subfigure}
    \caption{Real-time GPS positions of approximately 55 buses from the Mahanagar Yatayat Samiti, a metropolitan transport committee, as taken from the Mahanagar app. }        
        \label{fig:mahanagarAppShots}
\end{figure*}

\begin{figure}[h]
    \centering
    \includegraphics[width=0.6\textwidth]{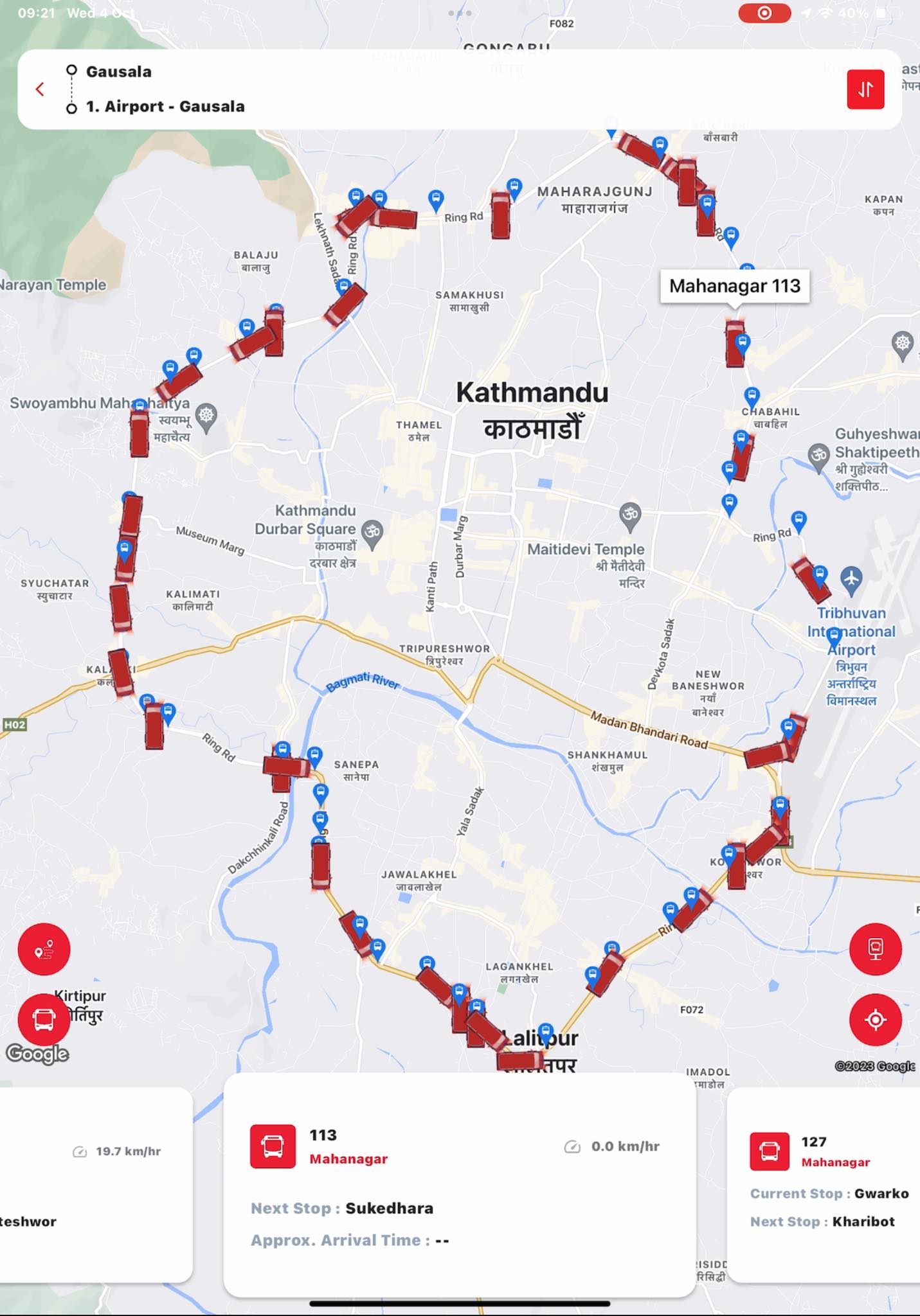}
    \caption{GPS position of buses at 9:21 AM, which shows an example of approximately uniform spacing among buses. }
    \label{fig:uniform}
\end{figure}

Mahanagar Yatayat Samiti, a metropolitan transport committee, operates approximately 55 buses along the Ring Road route.  However, there are numerous additional buses from different committees on the same route. It should be noted that these bus drivers still operate independently without adhering to a fixed timetable and thus cannot be modelled by standard bus route models following Poisson distribution. A significant development by the committee is the Mahanagar app, a smartphone application that enables users to track the GPS positions of each bus. In figure \ref{fig:mahanagarAppShots}, we have included time snapshots taken from the Mahangar app. These snapshots contain the location of buses run by the committee at various time stamps. In certain timestamps, such as in figure \ref{fig:uniform}, the spacing between buses can be seen to be uniform as well. This kind of information, taken from the app, can be leveraged for optimization purposes. 

\subsection{What is to optimize?}
The optimal spacing between buses can vary depending on whether you consider the perspective of bus drivers or passengers. From the perspective of a passenger waiting at a bus stop, the optimal spacing between buses would be when the spacing has a probability distribution close to the Poisson distribution. This type of distribution indicates a higher probability of buses arriving at closer intervals, increasing the likelihood of frequent encounters between buses. While this is better from the passenger's perspective,  it is in conflict with the effort of the drivers to maximize the number of transported passengers and, accordingly, maximize the distance to the preceding bus. Therefore, from the perspective of the driver,  the optimal probability distribution for spacing between buses would be one that closely approximates the Wigner-Dyson distribution. The third possibility would be a balance between these two cases in which the probability distribution would be close to the Brody distribution. Even though we have not provided proof that these are indeed optimal scenarios, for now, we leave it as a conjecture.

In this paper, we will mostly focus on the bus perspective of a driver who is trying to maximize his profit while following a particular route. Somewhat counter-intuitively, the optimal bus spacing distribution, in this case, would be when the system is chaotic. In the next section, we will provide techniques for such optimization procedures. The concept behind this approach involves utilizing the real-time position data obtained from GPS systems installed in buses. By employing a combination of quantum and classical algorithms and simulations, we then generate recommendations for the optimal stopping time at upcoming or current bus stops. These algorithms leverage the power of quantum computing alongside traditional computing techniques to process and analyze the bus position data, ultimately suggesting the most efficient and effective stopping times for bus operations.

\section{Optimizing the Ring road route}
\label{sec:optimization}
In this section, we will provide various quantum techniques to optimize the ring road route. To simplify the model, we assume that buses travel in a circular path, which is topologically equivalent to the ring road route. Therefore, the principles of physics and optimization can be applied similarly.

Let us denote the positions of N buses moving in a circle as $x_1, x_2, ..., x_N$, with the periodic boundary condition that $x_{N+1} = x_1$.  Let $G(x_1, x_2, ..., x_N)$ be the level spacing distribution for these positions. In the ring road route, these real-time positions of buses are given by GPS installed in buses. Now, the optimization is to provide new positions $x_i$ such that the new level spacing distribution is close to that of the Wigner-Dyson distribution for GUE, which we denote by $P_{\text{WD}}$. More concretely, if $x_1, x_2, ..., x_N$ is the position of the buses at time $t$, the task is to provide new bus positions $x_1 + \delta x_1, x_2 + \delta x_2, ...., x_N + \delta x_N $ which each bus should shift in time $t + \delta t$ such that 

$|G(x_1 + \delta x_1, x_2 + \delta x_2, ...., x_N + \delta x_N) - P_{\text{WD}}| \leq \epsilon$ 

There are two possible ways that one can shift the buses to  the distribution $G(x_i)$ to $P_{WD}$:
\begin{itemize}
    \item Decrease or increase the speed of the buses depending on the sign of $\delta x_i$
    \item Manipulate stopping time at the next stop depending on $\delta x$
\end{itemize}

However, in a busy road system with speed limits on highways, it might not be possible to increase the speed of the buses. This leaves the option to manipulate the stopping time of the next bus stop. To work out the algorithm for it, we need to introduce a few relevant terms:
\begin{enumerate}
    \item Due to the non-uniform speed of the buses and the chaotic nature of traffic, we would consider an average velocity between two stops $i$ and $j$ as $\braket{v_{ij}}$ where $i<j$.
    \item Since a bus stops at a bus stop for a considerable amount of time every time, we introduce a maximum allowed stopping time at each stop as $t_m$ and an average stopping time at bus stops as $\braket{t_s}$. The maximum stopping time $t_m$ may be due to city regulations that impose such limitations, or it could be motivated by the practical consideration that excessively long stopping times would allow buses from behind to catch up. 
\end{enumerate}

Depending on $\delta x$, a bus might need to stop at a bus stop for a longer period of time or for a shorter period of time. To consider all the scenarios, we considered three possible states:
\paragraph{Case I: $\delta x_i = 0$}
If $\delta x_i = 0$ for a bus indexed $i$, then the level spacing distribution resembles the WD distribution. The bus can stop at the next bus stop for an average time $\braket{t_s}$.
\paragraph{Case II: $\delta x_i \neq 0$}
A positive value of $\delta x_i$ means that the bus is lagging behind, considering the position given by the new optimized distribution, and the negative value means the bus is faster than the given position.  Consequently, the bus must stop at the bus stop for less or more time than $\braket{t_s}$ respectively. The bus would need to stop at the next bus stop for $ t= \braket{t_s} \pm \frac{\delta x_i}{\braket{v_{ij}}}$, where $j$ is the position of the next bus stop. 

However, when $\delta x_i < 0$, a bus cannot stop more than the maximum stopping time $t_m$. If $t\geq t_m$, the bus must stop at the bus stop for time $t_m$ and the remaining distance needed to lag, $d = \braket{v_{ij}}(\braket{t_s} - t_m) $ will be reconciled at the next bus stop.

% Written by Kiran Sir: It might be possible that some of the buses are at bus stops. In order to accurately model the presence of buses at bus stops, we will introduce a maximum allowed stopping time at each stop, denoted as $t_s$. This constraint may be due to city regulations that impose such limitations, or it could be motivated by the practical consideration that excessively long stopping times would allow buses from behind to catch up. Let $\langle v \rangle$ denote the mean velocity of buses along the route. Under this constraint, if $ \delta x_i / \langle v \rangle \leq t_s$, then $\delta x_i / \langle v \rangle$ would represent the stopping time at the bus stop. However, if $\delta x_i / \langle v \rangle \geq t_s$, the bus would stop for the maximum allowed time $t_s$ and then continue its journey, shifting by $ \delta x_i - \langle v \rangle t_s$. Obviously, this simple optimization can be done in a classical computer, too; however, for a more complicated Hamiltonian, we will need a quantum computer. 

\subsection{Spectral analysis}
We start with the Hamiltonian,
\begin{equation}
\label{eq:hamiltonianSpectral}
    H = \sum_{i = 1}^N \frac{m_i v_i^2}{2} + \sum_{i = 1}^N V(| x_{i-1} - x_i|) = \sum_{i = 1}^N H_i
\end{equation}
with $x_i$, $m_i$, and $v_i$ being the location, mass, and velocity of the $i^{th}$ vehicle with some guess for the potential function $U(| x_{i-1} - x_i|)$. It could just be $| x_{i-1} - x_i|$ as a starting case. Changing the parameters $x_i$ by $\delta x_i$ gives a new Hamiltonian $H'$. One can then check the eigenspectrum of this Hamiltonian using quantum computer \cite{Zisling_2021}. If the level spacing distribution is close to the Wigner-Dyson distribution, we will accept this new position of the buses. Otherwise, we will keep changing the $\delta x_i$. We do this until we reach the level spacing distribution close to the Wigner-Dyson distribution. When this happens, the Hamiltonian is expected to be close to the Dyson gas Hamiltonian with potential,
\begin{equation}
\label{eq:Dyson_gas}
    V = - \sum_{i < j} \log (|x_i - x_j|)
\end{equation}
One can obtain the new optimal parameters for bus positions by solving this. In addition to the level spacing distribution analysis, we can use additional diagnostics, using quantum computer \cite{Joshi_2022}, to check if chaos has emerged. 
Given the level spacings $S_i = \lambda_{i+1} - \lambda_i$, we can define the following ratios:
\begin{equation}
    r_i = \frac{\text{Min}(S_i, S_{i+1})}{\text{Max}(S_i, S_{i+1})}
\end{equation}
For Poisson-like level spacing distribution, $r_i \approx 0.38629$, while for Wigner-Dyson-like level spacing distribution, $r_i \approx 0.60266$. The obtained $r_i$ should be close to $0.60266$ to get the optimal bus spacing. In this scheme, the task is reduced to obtaining the eigenvalues of the Hamiltonian using a quantum computer, analyzing it in classical machines, and updating the Hamiltonian.

One approach to achieve this goal involves conducting a quantum simulation of the Hamiltonian. One can analyze its eigenvalues spectrum by manipulating the Hamiltonian parameters. Alternatively, this simulation can be directly performed on a quantum computer \cite{nielsen2001quantum}. In this method, we are provided with a black box/oracle that has simulated unitary  \(U = e^{-iHt}\) and powers of it. This step might be costly and is usually considered given in the algorithm.  Utilizing quantum phase estimation, eigenvalues can be obtained with an oracle complexity of \(O(n^2)\), where \(O\) represents Big O notation, which represents the upper bound of the growth of an algorithm as its input size grows, and \(n\) is the bit length or number of qubits \cite{nielsen2001quantum}. Similar to the variational technique, one has the flexibility to adjust both the parameters within the Hamiltonian and the unitary operator, allowing for a more adaptable and efficient execution of the algorithm.

\subsection{Output wave function analysis}
In the previous section, we considered analyzing the output spectrum for the optimization process. However, the technique described therein involved complicated quantum circuits, like quantum phase estimation with significant depth, making it difficult to implement in the near term. One could also directly analyze the output wave function generated by a parameterized quantum circuit, which can be performed efficiently within polynomial time \cite{yang2023phasesensitive,lerma2019dynamical}.

One potential metric for analysis is the survival probability (\(\text{SP}\)), defined as the overlap between a given initial state \(\ket{\psi}\) and the output quantum state \(\ket{\phi}\):

\[
\text{SP} = |\langle \psi | \phi \rangle|^2
\]
This quantity indicates the likelihood of finding the initial system in its original state even after undergoing a unitary transformation. Employing the Hadamard test, one can calculate \(\text{SP}\). A value of \(1\) signifies that the output state closely matches the initial state, while \(0\) indicates complete orthogonality. In the context of a chaotic quantum system, one would expect \(\text{SP}\) to be close to \(0\). Consequently, the Hamiltonian is adjusted iteratively until \(\text{SP}\) decreases, and the corresponding Hamiltonian parameters are noted.

It is crucial to note that these parameters might not necessarily be optimal. Achieving true optimality would necessitate a comprehensive spectral analysis, a task demanding powerful quantum computers. Nevertheless, valuable insights can still be extracted from the partial information gathered, guiding the optimization process. Additionally, alternative metrics such as inverse participation ratio spread complexity and Out-of-Time-Order correlation experiments \cite{Green_2022} can also be employed for analyzing the output wave function.

\section{Outlook}
\label{sec:outlook}
Our research aims to state the conditions in which we can claim that the Kathmandu bus system is indeed chaotic via the exploration of quantum signatures. To identify these signatures, it is essential to gather comprehensive data on the system, including bus positions, passenger loads, traffic conditions, schedules, and other relevant parameters. The exploration of bus systems as coherent chaotic systems in this study opens the door to more detailed traffic flow analysis. Researchers can delve into the dynamics of traffic, which display critical points, non-equilibrium phase transitions, noise-induced transitions, and fluctuation-induced ordering phenomena. By adopting a quantum many-body system perspective, researchers can uncover chaotic behaviour in the interaction of complex traffic patterns and contribute to the field of traffic science.

Another research direction is to validate the proposed quantum optimization paradigms using real data. By applying classical and quantum optimization algorithms to this real data, a direct comparison can be made to assess the effectiveness of both approaches in improving bus system performance. Of course, it is not feasible to actually run the algorithm on a quantum computer due to the severe limitations of quantum computers today. However, the algorithm we suggest is to optimize the bus system to achieve Wigner-Dyson distribution, which is observed a lot in quantum systems. This quantum-inspired algorithm can be tested against classical optimization techniques and it is interesting to compare the differences between them. 

Besides, future research can also take a holistic approach by considering optimization from the perspective of transportation authorities as well who seek to minimize congestion and environmental impact unlike passengers who may value shorter waiting times and drivers who aim to maximize profit. It is interesting to explore multi-objective optimization via quantum route and to see if these perspectives can lead to inspirations for more robust and adaptable bus system designs.

In summary, future research in this area is promising and multi-faceted. Simulating Dyson gas, conducting in-depth traffic flow analysis, collecting real data for validation, and approaching optimization from various perspectives are all avenues that can further advance our understanding of bus systems and potentially lead to more efficient and sustainable urban transportation solutions.

%\begin{itemize}
%    \item simulates Dyson gas
%     \item Analysis for more detailed traffic flow. Consequently, traffic displays critical points, non-equilibrium phase transitions, noise-induced transitions, and fluctuation-induced ordering phenomena, which are very interesting to study from quantum many-body system perspective 
%     \item Collect real data of Kathmandu bus system, and compare classical vs quantum optimization paradigm
%   \item optimization from all three prospective
  
% \end{itemize}
\section{Conclusion}
\label{sec:conclusion}
In an endeavour to optimize the transportation system along Kathmandu's Ring Road, this study has investigated a novel approach integrating principles from quantum chaos and Random Matrix theory. Ring Road of Kathmandu serves as a perfect model for optimization due to its unique, closed-loop design, which allows for a simplified model of buses moving in a one-dimensional gas format, enabling numerical and quantum simulations. The road sees approximately 55 buses from the Mahanagar Yatayat Samiti, among other vehicles, which often operate without a fixed timetable. The study explored three optimization scenarios: one from the perspective of the passenger, another from the bus view of the driver, and the third being a balance between the two. From the standpoint of a passenger, ideal bus intervals would follow a Poisson distribution, ensuring frequent bus arrivals. For drivers, the Wigner-Dyson distribution was deemed optimal as they aim to maximize the distance to the preceding bus. The paper primarily leaned toward the bus perspective of the driver. A smartphone application called the Mahanagar app, which tracks the GPS locations of the buses in real time, was used to study the properties of the bus system. The suggested technique involves adjusting the positions of the buses so that their spacing closely aligns with the desired Wigner-Dyson distribution. Two practical methods to achieve the desired spacing were proposed: manipulating the bus speeds or adjusting their stopping times at the bus stops. Given road conditions and regulatory restrictions, the latter was deemed more feasible. Quantum algorithms, combined with classical methods, were suggested to analyze bus position data and generate recommendations for optimal stopping times. The Hamiltonian was shown to be able to be formulated with variables representing the positions of vehicles or their spacing. By adjusting the parameters of the Hamiltonian, the study aimed to bring its eigenspectrum closer to the Wigner-Dyson distribution. A secondary method was proposed for scenarios where deep quantum circuits, like those in quantum phase estimation, are infeasible. This involves analyzing the output wave function of a parameterized circuit and computing metrics like the survival probability using the SWAP test. This provides insights into how chaotic the quantum system is, which could help in the optimization process. In conclusion, by suggesting the application of quantum principles with real-world traffic data, this paper has paved the way for innovative solutions to optimize the flow of buses on Kathmandu's Ring Road. While full implementation requires powerful quantum computers, the insights garnered from this approach can provide significant improvements to the current transport system.

\section{Acknowledgements}
This research is part of the Munich Quantum Valley, which is supported by the Bavarian state government with funds from the Hightech Agenda Bayern Plus. We also acknowledge support from 
Physics Without Frontiers (PWF) program of the International Centre for Theoretical Physics (ICTP), Italy.

\bibliographystyle{splncs04}
\bibliography{main}


\end{document}